# Uniform Graphene on Cu and Au Substrates

Ch. Ravi Prakash Patel[a*], K. Awasthi[b] and Thaur Prasad Yadav[a]

[a]Hydrogen Energy Centre, Department of Physics, Institute of Science,
Banaras Hindu University, Varanasi-221005, India
[b] KN Government PG College Gyanpur, Bhadohi 221304, India
[*]Corresponding author. E-mail: ravi2008prakash@gmail.com

**ABSTRACT**
We report the synthesis of single and bi-layer graphene films by low pressure chemical vapor deposition (LPCVD) technique on Cu and Au substrates. The as grown films were characterized by transmission electron microscopy (TEM), scanning electron microscopy (SEM) and Raman spectroscopy techniques. As evidence by SEM and TEM, the growth takes place through surface adsorption. The large lateral area (1cm x 1cm) graphene deposited on Cu can easily be transferred on $Si/SiO_2$. It has been shown that in Cu Substrate, it is adsorption of carbon on the substrate through which graphene gets formed. In the case of Au substrate both the adsorption and diffusion-precipitation leads to the growth of graphene.

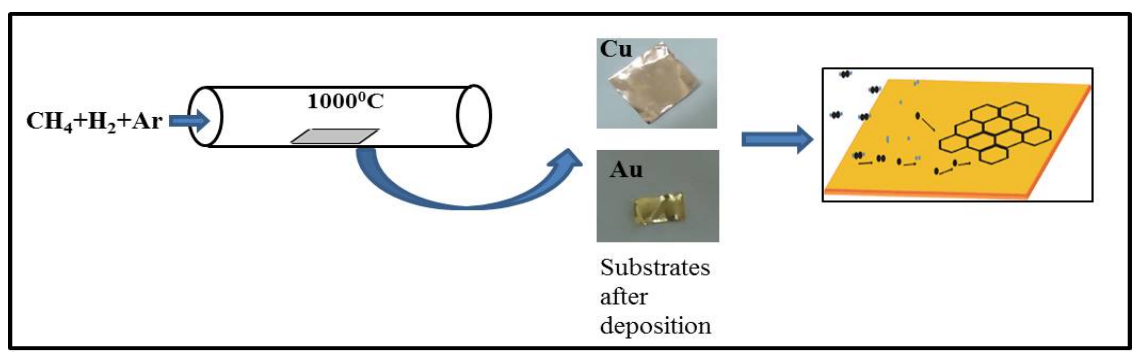

## 1. Introduction

Graphene is one of the most advanced materials due to fascinating properties such as high carrier mobility, high electrical conductivity, as well as excellent mechanical flexibility and strength [1,2] .Many unique physical and chemical properties of graphene making it a promising applications in various fields such as batteries, supercapacitors, fuel cells, optoelectronics, photovoltaic devices, and biosensors [2-4]. The synthesis of large-area high-quality graphene is required for utilizing these potential applications.There are several methods used to synthesize graphene films such as mechanical cleavage of highly oriented pyrolytic graphite [6],sublimation of epitaxial SiC [7], chemical reduction of graphene oxide [8] and catalyst-assisted chemical vapour deposition (CVD [9,10]. Obtaining single layer graphene (SLG) and few layer graphene (FLG) via mechanical cleavage method is not practical due to the small flake size and non-uniformity in the number of graphene layer in the flake exfoliated from graphite. Also this methods is time taking. High mobility (~10000 $cm^2$/v-s) epitaxial graphene can be obtained through thermal decomposition of SiC [11]. However, high cost and limited SiC wafer size may hold back its broad applications. Graphene synthesized by chemical reduction of graphite oxide (GO) has high yield, but major disadvantage is low electrical mobility originating from their defective structures. Beside these methods, metal catalysts based chemical vapor deposition (CVD) method is a most attractive ways for synthesizing large area and high quality graphene.The different metals such as nickel, ruthenium, iridium, platinum, palladium, and cobalt have used as a substrate for the growth of graphene [9, 12-16]. In few report, the Cu and Au have been used as substrates for synthesis of graphene by CVD method [12, 17-19]. The use of Cu film as a catalyst was demonstrated for growth of single/bilayer graphene. In the present investigation, we describe the growth of graphene film on Cu and Au substrates using LPCVD method. A comparison of the graphene deposits has also been made.

## 2. Experimental
### 2.1 Experimental setup
A low pressure chemical vapor deposition (LPCVD) (Etamota, USA) system with 5 cm diameter and one meter length quartz tube has been used in the present investigation. The LPCVD system has three heating zones furnace, which can be set at different temperatures. Fig. 1 shows schematic diagram of LPCVD system, where three substrates can be placed in each experimental run at different heating zone. The $S_1$, $S_2$, $S_3$, $S_4$ and $S_5$ are valves transmission of specific gasses ethane, hydrogen ($H_2$), Methane ($CH_4$), Argon (Ar) and liquid vapor controller (Bubbler) respectively. The LPCVD furnace is connected to vacuum pump with two vacuum gauges and a butter fly valve.

### 2.2 Synthesis of Graphene
The Cu and Au (Alfa Aesar, 99.99% pure) 25 μm thick (1 cm x 1 cm) were used as a substrates. The Cu and Au substrates were sonicated in acetone for 30 minutes and then inserted inside the quartz tube of LPCVD system. The chamber was evacuated and heated in the presence of 15% $H_2$ mixed argon gas at 150 torr pressure. The temperatures employed were in the range of 900$^o$C to 1000$^o$C. After annealing the substrates, the gases methane ($CH_4$) (10sccm), $H_2$ (10sccm) and Ar (20sccm) were flown in the chamber. The optimized synthesis temperature for Cu was 1000$^o$C and for Au 900$^o$C. The graphene film was grown at Cu for time duration of 10 and 12 minutes and for Au, 6 and 10 minutes. Finally, fast cooling was performed by opening the split furnace in the presence of argon at 150 torr pressure.

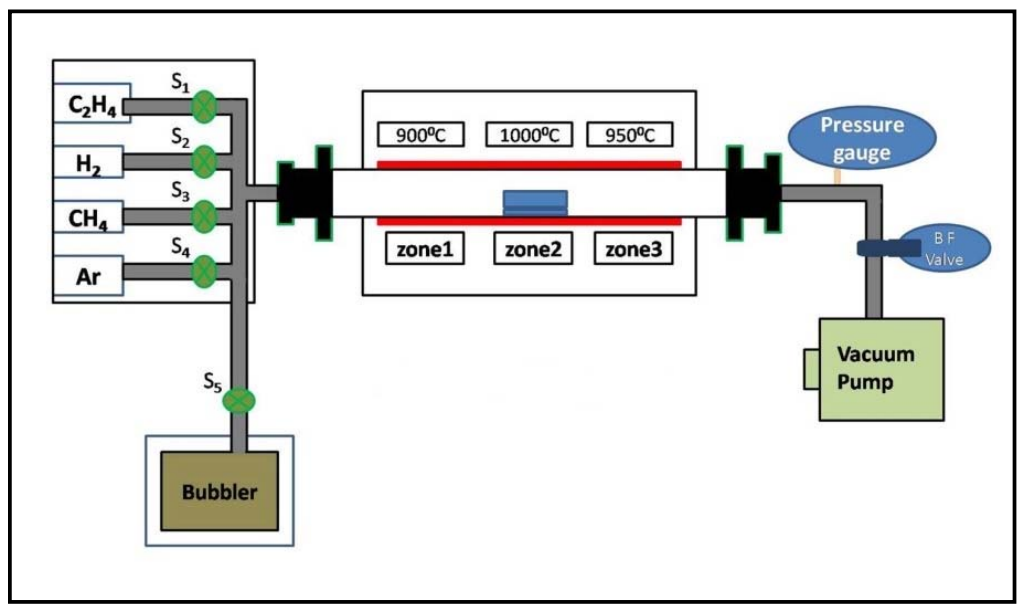

**Fig. 1**: The schematic diagram of our LPCVD system.

*2.3. Characterization techniques*

The structural characterization of the as-grown material have been carried out using X-ray diffraction (XRD) employing X'Pert PRO PANalyticaldiffractometer equipped with a graphite Monochromator with a Cu source ($\lambda = 1.5402$Å, $CuK_\alpha$ operating at 45kV 40mA). The microstructural characterizations were carried out using a scanning electron microscope (SEM: QUANTA 200). The detailed structural and microstructural characterization was carried out using transmission electron microscopy (TEM) (TECNAI-20$G^2$ at 200kV) in the diffraction and imaging modes. Raman spectra were recorded from Horiba JobinYuon (HR 800) Raman Spectrometer model no. H 45517 using an argon ion laser source $\lambda = 514.532$nm.

*2.4. Transfer of graphene films*

As grown graphene on the Cu and Au foil has been transferred on Si/$SiO_2$ substrate for the characterization. The as grown graphene was transferred on $SiO_2$ substrate as follows: poly (methyl methacrylate) (PMMA) was dissolved in choloro-benzene in the ratio of 50mg/ml and then the resulting solution was spin coated on the as grown graphene (~ 200nm thin layer) at 5000 rpm. Then it was dried under vacuum for 2 hours. The Cu and Au foils were then dissolved in dilute $HNO_3$ and $HNO_3$-HCl solutions, respectively. The substrate (Cu/Au) gets dissolved leaving behind graphene stacked with PMMA. The stacked PMMA-graphene film was washed with distilled water for 4 to 5 times, so that the acids gets removed. The stacked PMMA-graphene was transferred by floating it on distilled water and then lifting it on Si/$SiO_2$ substrate. This was then dried under 100 watt electric bulb. The graphene film sits on $SiO_2$ above the Si surface. After drying, the PMMA film was immersed in acetone. The PMMA is dissolved in acetone and the graphene film remains on $SiO_2$ substrate.

**3. Results and discussion**

To explore the microstructure of Cu and Au substrates / foils, XRD pattern before and after annealing were recorded. Fig. 2 show representative XRD patterns taken from Au substrate / foil before and after annealing. All the peaks were indexed very nicely by Au with lattice parameter a= 4.079Å. The most significant difference between the two photographs is that on annealing the (111) peak becomes mere dominant as compared to the corresponding peak in the un-annealed. Au substrate: This implies that the lattice plane on which graphene grows as (111). It may be noted that this is also the close packed plane of atoms for FCC gold lattice structure. A similar behavior relating to dominance of (111) plane on annealing has been observed for Cu substrate / foil also. This is expected since for both Cu and Au lattice structure is FCC.The surface morphology of graphene grown on substrates Cu and Au has been carried out by SEM Fig. 3(a) shows typical SEM image of annealed Cu foil. Fig. 3(b) is the SEM image of the graphene grown on annealed Cu foil. As can be seen the microstructure of the annealed Cu substrate reveals the polycrystalline nature. Several grain boundaries can be easily noticed. For the Cu substrate with deposited graphene film, the microstructure still shows grain boundaries of the Cu substrate. But the contrast has become dimmed suggesting the presence of graphene layer. This type of SEM characteristics was observed for several graphene films deposited on Cu substrate. The representative SEM micrograph of graphene film on Au substrate is shown in Fig. 3(c). Here again the presence of graphene can be inferred by the dimmed contrast of the underlying grain structure of Au substrate. The presence of graphene layer of Cu / Au substrate was further confirmed by Raman Spectroscopy to be described in the next section. One clear cut difference between the microstructures of graphene on Cu and Au is the presence of precipitate like islands on the graphene film grown on Au substrates. These precipitate islands were invariably present for all the graphene films grown on Au. Some of these islands are marked by arrows.

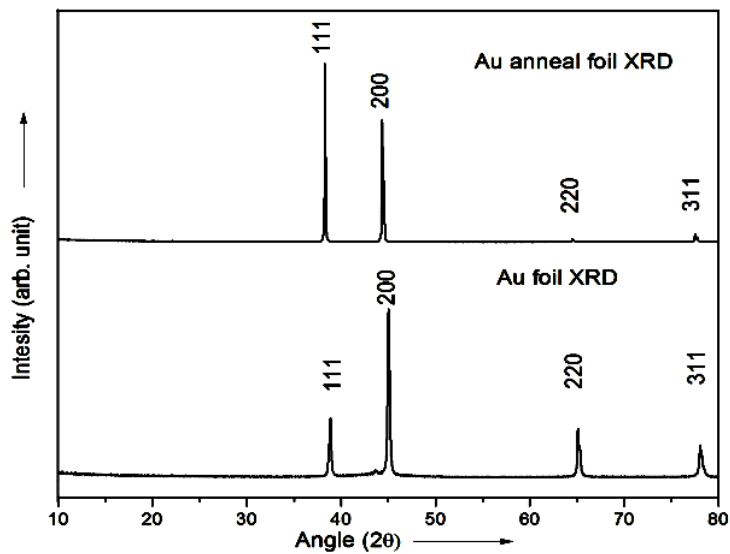

**Fig. 2:** XRD pattern of the Au substrates before annealing and after annealing.

The as grown graphene films were transferred on the Cu grids for TEM characterization. To determine the lateral size of as growngraphene, low magnification images were taken. Fig. 4(a) shows the large and uniform graphene film grown by LPCVD method. The inset of Fig. 4(a) is the selected area diffraction pattern of the as grown graphene. This show the six fold symmetrical distribution of diffraction spots (with d ~3.35 Å and a ~2.464 Å). This is the characteristic of graphene. In order to confirm the number of layers in the as synthesized graphene, high resolution TEM micrograph was employed. Fig. 4(b) and 4(c)exhibit the high resolution TEM image of graphene transferred from Cu foil. There do not seem to be lattice fringes suggesting multilayer graphene. Thus the as deposited graphene can be taken to be single (or few layer) graphene.

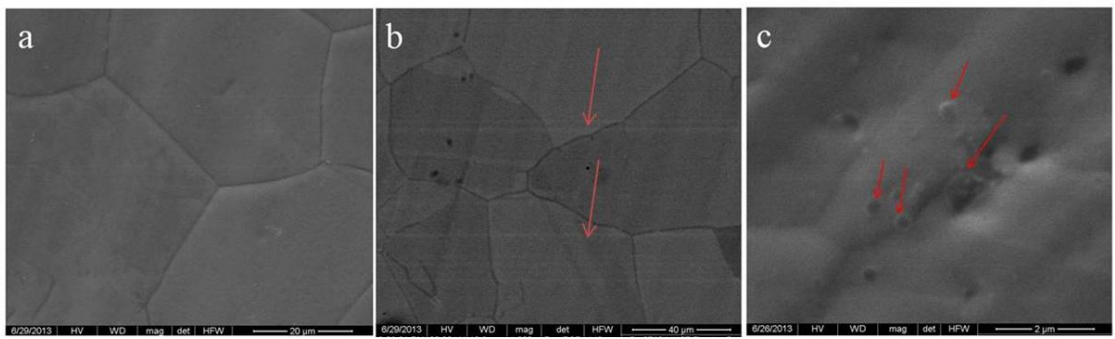

**Fig. 3:** SEM images of annealed Cu foil (a), graphene growth of Cu substrate (b)and graphene growth of Au substrate (c).

Raman spectroscopy is very powerful tool for finger printing graphene. Fig. 5 shows the Raman spectra of graphene substrates after transfer of graphene on the $SiO_2$ substrate. Raman spectra for graphene transferred from Cu, clearly shows (Fig. 5 (a)) G and 2D bands at 1589 and 2704 $cm^{-1}$ for the graphene grown in time duration of 12 minutes. For graphene (grown for 10 minutes) transferred from Cu substrate bands are at 1583 and 2688 $cm^{-1}$ as shown in Fig. 5(b). The $I_{2D}/I_G$ ratio is 0.97 and 3.96 for Fig. 5 (a) & (b), respectively. On this basis of these results coupled with known facts about Raman peaks in graphene, it can be said that the graphene represented by Fig. 5 (b) is single layer graphene (SLG) and that by 5 (a) is bi-layer graphene (BLG). Fig. 5(c) and (d) is the Raman spectra of graphene transferred from Au. The peaks position of D, G & 2D are at 1351, 1573 and 2694 cm$^{-1}$, respectively for graphene (grown in 10 minutes in Fig. 5 (c)). For graphene (grown in 6 minutes) the peak position of G and 2D bands are 1656 and 2684 cm$^{-1}$ respectively. The $I_{2D}/I_G$ ratio calculated for Fig. 5(c) and 5(d) are 1.1 and 4.06 respectively. Based on the above analysis of Raman spectra for graphene, Fig. 5 (d) represents single layer (SLG) and Fig. 5 (c) as bi-layer (BLG) graphene.

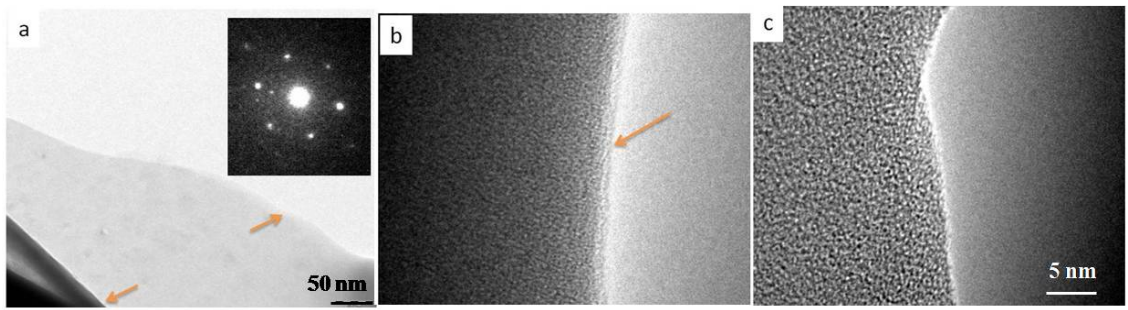

**Fig. 4:** (a) Low magnification TEM image of transferred graphene film on Cu grid from Cu substrate. (b) and (c) high resolution TEM images of graphene films grown using Cu and Au substrates, respectively.

Two different mechanisms have been suggested for the growth of graphene on metallic substrates [17-19]. One is based on diffusion and precipitation of carbon from hydrocarbon source (like CH$_4$) on the metal substrate. The other growth mechanism takes recourse to adsorption of carbon on the metal (substrate) surface. Islands of monolayer carbon get formed on the metal surface. Additional carbon atoms arriving at the surface see metal atoms only around the previously grown carbon islands. Further adsorption of carbon atoms then takes place on these metal atoms leading to further growth of carbon monolayer. This eventually leads to formation of graphene layer. The results of growth of graphene on Cu and Au substrates investigated in the present study was analyzed in the light of the above two mechanisms. The solubility of carbon is low in both Cu and Au, it appears that the adsorption mechanisms plays dominant role in the growth of graphene layer for both Cu and Au.

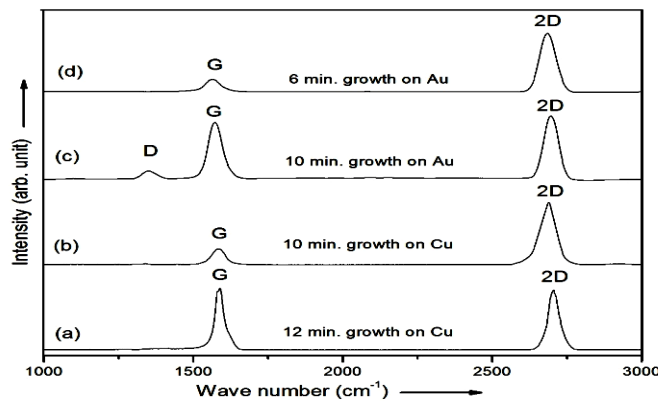

**Fig. 5:** Raman spectra of transferred graphene film on SiO$_2$ substrate (a) & (b) from Cu substrate and (c) & (d) from Au substrate

However, it should be mentioned that the graphene layer grown on Au substrate invariably shows the presence of carbon precipitate islands (Fig. 3 (c)). This suggests that for Au diffusion and precipitation of carbon also plays role in the growth of graphene layers. It can therefore be said that whereas for Cu substrate the graphene film grows only by surface adsorption of carbon atoms, however for Au substrate besides surface adsorption, diffusion and precipitation also plays a role in the growth of graphene layers. This observation is supported by the fact that the solubility of carbon in Au is 0.06% whereas for Cu it is 0.04%. The higher solubility of Carbon in Au can trigger the growth mechanism involving diffusion and precipitation. This result is different than those obtained by other workers [20-22].

## 4. Conclusion

In the present investigation we have successfully grown single and bi-layer graphene by LPCVD method. Cu and Au were used as a substrate which acted as a catalyst for graphene growth. The grown graphene were transferred on $SiO_2$ and were characterized through XRD, SEM, TEM and Raman spectroscopy. These studies have led to two important results. One relates to the fact that single and bi-layer graphene can be grown by LPCVD on Cu and Au substrates. The other results are for the growth of graphene on Cu and Au. On Cu the growth talks place through adsorption of carbon atoms. On the other hand on Au substrate both; adsorption and diffusion cum precipitation leads to growth of graphene.

## Acknowledgements
We would like to dedicate this article in the memory of my beloved and respected teacher and Ph.D. supervisor (Prof. O.N. Srivastava) who passed away due to COVID-19 ailment.